\documentclass[aps,prl,reprint,superscriptaddress]{revtex4-1}

\newcommand{\einh}[1]{\ensuremath{\,\text{#1}}}
\newcommand{\etal}{\textit{et al}}

\usepackage{mathptmx}

\usepackage{upgreek}
\usepackage{graphicx}
\usepackage[colorlinks, citecolor=blue, linkcolor=blue, urlcolor=blue]{hyperref}

\usepackage[alsoload={physical}, per=slash]{siunitx}
\newunit{\atomic}{\textmd{a{.}u{.}}}
\begin{document}

%\preprint{}

\title{Time-Resolved Measurement of Interatomic Coulombic Decay in $\text{Ne}_{2}$}



%__________________________________-------------------------------------------------
%
\author{K.~Schnorr}
\email[]{Kirsten.Schnorr@mpi-hd.mpg.de}
\author{A.~Senftleben}
\author{M.~Kurka}
\affiliation{Max-Planck-Institut f\"ur Kernphysik, 69117 Heidelberg, Germany}

\author{A.~Rudenko}
\affiliation{Max-Planck Advanced Study Group at CFEL, 22607 Hamburg, Germany} 

\affiliation{J.R. Macdonald Laboratory, Kansas State University, Manhattan, Kansas 66506, USA}

\author{L.~Foucar}
\affiliation{Max-Planck Advanced Study Group at CFEL, 22607 Hamburg, Germany} 
\affiliation{Max-Planck-Institut f\"ur medizinische Forschung, 69120 Heidelberg, Germany}

\author{G.~Schmid}
\author{A.~Broska}
\author{T.~Pfeifer}
\author{K.~Meyer}
\affiliation{Max-Planck-Institut f\"ur Kernphysik, 69117 Heidelberg, Germany}

\author{D.~Anielski}
\author{R.~Boll}
\affiliation{Max-Planck-Institut f\"ur Kernphysik, 69117 Heidelberg, Germany}
\affiliation{Max-Planck Advanced Study Group at CFEL, 22607 Hamburg, Germany} 

\author{D.~Rolles}
\affiliation{Max-Planck Advanced Study Group at CFEL, 22607 Hamburg, Germany} 
\affiliation{Deutsches Elektronen-Synchrotron, 22607 Hamburg, Germany}

\author{M.~K\"ubel}
\affiliation{Max-Planck-Institut f\"ur Quantenoptik, 85748 Garching, Germany}

\author{M.F.~Kling}
\affiliation{J.R. Macdonald Laboratory, Kansas State University, Manhattan, Kansas 66506, USA}
\affiliation{Max-Planck-Institut f\"ur Quantenoptik, 85748 Garching, Germany}

\author{Y.H.~Jiang}
\affiliation{Shanghai Advanced Research Institute, Chinese Academy of Sciences, Shanghai 201210, China}

\author{S.~Mondal}
\author{T.~Tachibana}
\author{K.~Ueda}
\affiliation{Institute of Multidisciplinary Research for Advanced Materials, Tohoku University, Sendai 980-8577, Japan}

\author{T.~Marchenko}
\author{M.~Simon}
\affiliation{Laboratoire de Chimie Physique-Mati\`ere et Rayonnement, UPMC and CNRS, 75231 Paris, France}

\author{G.~Brenner}
\author{R.~Treusch}
\affiliation{Deutsches Elektronen-Synchrotron, 22607 Hamburg, Germany}

\author{S.~Scheit}
\affiliation{Goethe-Universit\"at, 60438 Frankfurt, Germany} 

\author{V.~Averbukh}
\affiliation{Imperial College London, London SW7 2AZ, United Kingdom}

\author{J.~Ullrich}
\affiliation{Physikalisch-Technische Bundesanstalt, 38116 Braunschweig, Germany}

\author{C.D.~Schr\"oter}
\author{R.~Moshammer}
\affiliation{Max-Planck-Institut f\"ur Kernphysik, 69117 Heidelberg, Germany}

%__________________________________-------------------------------------------------

%\thanks{}

\date{\today}

\begin{abstract}

The lifetime of interatomic Coulombic decay (ICD)  \cite{Cederbaum97} in $\text{Ne}_{2}$ is determined via an extreme ultraviolet pump-probe experiment at the Free-Electron Laser in Hamburg. The pump pulse creates a $2s$ inner-shell vacancy in one of the two Ne atoms, whereupon the ionized dimer undergoes ICD resulting in a repulsive $\text{Ne}^{+}(2p^{-1}) - \text{Ne}^{+}(2p^{-1})$ state, which is probed with a second pulse, removing a further electron. The yield of coincident $\text{Ne}^{+} - \text{Ne}^{2+}$ pairs is recorded as a function of the pump-probe delay, allowing us to deduce the ICD lifetime of the  $\text{Ne}_{2}^{+}(2s^{-1})$ state to be $(150 \pm 50) \einh{fs}$ in agreement with quantum calculations.

\end{abstract}

%\keywords{}

\maketitle

Interatomic or intermolecular Coulombic decay (ICD), a radiationless relaxation process resulting from long-range correlation effects, was proposed theoretically by Cederbaum~\etal. in 1997~\cite{Cederbaum97}:
An inner-shell vacancy is filled by an outer-valence electron, while the released energy is transferred to a neighboring atom, which emits an electron into the continuum. The first experimental observation of ICD was reported on neon clusters \cite{Marburger03}, while Jahnke \etal. performed an unambiguous experiment on neon dimers \cite{Jahnke04}. One of the most remarkable features of ICD is the extremely short lifetime, making it a highly efficient decay process for an excited atom embedded in an environment and, thus, a strong source of low-energy electrons, which for instance, cause damage in tissue in radiation therapy \cite{DNA}.

Theoretically predicted ICD lifetimes span over several orders of magnitude depending on the system and the type of ICD mechanism. First calculations predicted a few, up to a hundred femtoseconds for neon clusters, depending on the cluster size \cite{Santra01_3}. The ICD lifetime of bulk atoms in large neon clusters of  $\SI{6(1)}{\fs}$ has been determined indirectly by measuring the width of the $2s$ photoelectron spectrum \cite{ohrwall}. Endohedral fullerene complexes are expected to be remarkably fast examples where a $2s$ vacancy in the $\text{Ne}^+$ ion, embedded in the fullerene, relaxes via ICD within $\SI{2}{\fs}$ \cite{ICDfast}. Further examples of a few-femtosecond ICD can be found in large water clusters \cite{Mucke10} as well as in $(\text{H}_{2}\text{O})_{2}$ \cite{Jahnke10}. For the latter ICD is faster than decay via proton transfer, which occurs within tens of femtoseconds. 

In the examples presented thus far, nuclear motion can be neglected as it typically takes place on time scales of hundreds of femtoseconds. However, this assumption does not hold for a slowly decaying system \cite{Jahnke07_2,ueda,Demekhin091,ICD_He,jahnke2013} giving it enough time to vibrate prior to its decay into a different electronic state. As ICD is a highly distance-dependent effect, nuclear motion has to be taken into account. If the overlap of the electronic orbitals between the interacting particles can be neglected, the decay width $\Gamma$ shows the characteristic dipole-dipole interaction behavior of $\Gamma \sim {R^{-6}}$. This assumption is reasonable for very weakly bound systems such as rare-gas clusters or other systems at large internuclear distances $R$ \cite{averbukh2004}. For small $R$ the enhanced orbital overlap can increase the decay width significantly \cite{averbukh2004}.

The ICD lifetime of the $2s$ vacancy in $\text{Ne}^{+}_{2}$ has been subject to several theoretical investigations. For a fixed internuclear distance of $\SI{3.2}{\angstrom}$, lifetime predictions of the intermediate $2^2 \Sigma _{u}^{+}$ state between ${64}$ and $\SI{92}{\fs}$ have been reported, depending on the applied method \cite{Santra01_2,fano-adc,Vaval07}. A fast decay time is consistent with the results from the first kinematically complete experiment on $\text{Ne}_{2}$, where a small broadening of the ions' kinetic energy was explained by only little nuclear motion, which corresponds to a fast ICD \cite{Jahnke04}. However, even a small change in $R$ influences the lifetime significantly \cite{Santra00_1} calling for calculations including nuclear motion \cite{Scheit03}.

Here, we present a direct measurement of the ICD lifetime in $\text{Ne}_{2}$ via a pump-probe experiment in the extreme ultraviolet (XUV) wavelengths regime. With this technique no precise spectroscopic \textit{a priori} knowledge of the potential-energy curves (PECs) or the transition rates is needed. Instead it can be extracted from the measurement. We realize the time-resolved experiment on ICD by combining the pump-probe technique with ion-ion coincidence spectroscopy. While the femtosecond XUV pulses at high intensities are delivered by the Free-Electron Laser (FEL) in Hamburg (FLASH), a reaction microscope is used for spectroscopy. 

The experiment has been carried out at the unfocused branch of beam line BL3 at FLASH with a FEL intensity of approximately $\SI{e12}{\watt\per\centi\meter\tothe{2}}$ for a single pulse -- chosen such that predominantly one-photon absorption in $\text{Ne}$ takes place -- at an average repetition rate of $800\, {\text{pulses}}/{\text{s}}$. The incoming FEL beam of $\SI{58.2}{\electronvolt}$ with a pulse duration of approximately $60 \einh{fs}$ (FWHM) is reflected off a multilayer split-and-delay mirror setup, which geometrically splits the beam into two halves with adjustable time delay $t_D$ and focuses them to a spot size of about $\SI{10}{\micro\meter}$ diameter into a supersonic gas jet. Ne dimers are produced by expanding Ne gas through a $\SI{30}{\micro\meter}$ diameter nozzle, kept at room temperature, with an injection pressure of $21 \einh{bar}$ ensuring negligible formation of larger clusters. The internal temperature of the gas jet was low enough to enforce virtually exclusive population of the vibrational ground state. The created ion pairs, emerging from Coulomb explosion after ionization of the dimers, are accelerated by a homogeneous electric field onto a time- and position-sensitive detector, which allows us to reconstruct their momentum vectors \cite{REMI}. 

\begin{figure}
\includegraphics[scale=0.72]{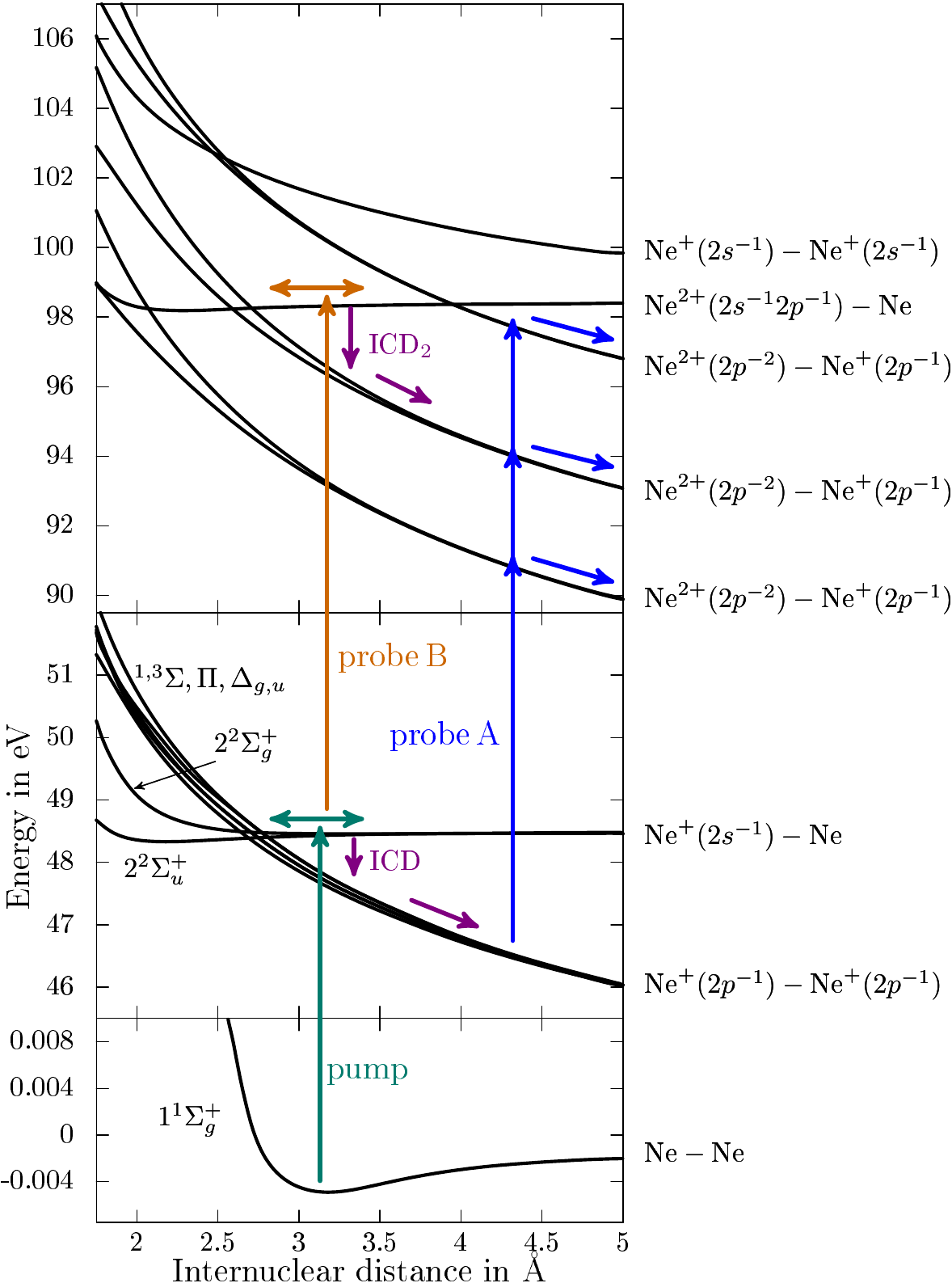}
\caption{\label{fig:potential} Relevant potential energy curves of $\text{Ne}_{2}$ \cite{Stoychev08_2} with possible pathways resulting in coincident $\text{Ne}^{+} - \text{Ne}^{2+}$ fragments, assuming single-photon absorption in both, the pump and the probe pulse.} 
\end{figure}

The kinetic energy release (KER) of both coincident ions equals to good approximation the energy of pointlike particles repelling each other according to the PECs shown in Fig.\,\ref{fig:potential}. The states can be populated by single photon absorption during the pump and/or probe pulse, or ICD after photoionization. Thus, by continuously scanning $t_D$ and recording the corresponding KER of each ion-ion coincidence, the nuclear dynamics is traced in real time. This way, we can map intermediate PECs, which are otherwise only accessible through theoretical calculations. Moreover, the delay-dependent yield of effectively triply charged dimers allows us to extract the ICD lifetime. 

During the measurement, the pump pulse creates a $2s$ inner-valence vacancy in one of the two Ne atoms and thereby initiates a nuclear wave packet in the $2^2 \Sigma _{g}^{+}$ or $2^2 \Sigma _{u}^{+}$ state of the $\text{Ne}^{+}_{2}$ molecular ion (Fig.\,\ref{fig:potential}). While the wave packet moves towards smaller internuclear distances in the attractive $2^2 \Sigma _{u}^{+}$ state, it essentially disperses on the flat $2^2 \Sigma _{g}^{+}$ potential, which has a very shallow minimum around $\SI{3.2}{\angstrom}$ \cite{Santra00_1}. The wave packet propagates until the system decays into one of the various repulsive $\text{Ne}^{+} (2p^{-1}) - \text{Ne}^{+} (2p^{-1})$ states via ICD. After the adjustable time delay $t_D$, the probe pulse removes another electron from one of the $\text{Ne}^{+}$ ions which results in a $\text{Ne}^{2+} (2p^{-2}) - \text{Ne}^{+} (2p^{-1})$ state (probe $A$ in Fig.\,\ref{fig:potential}). For the present moderate FEL intensity with dominant single photon absorption during the pulse, this effectively triply ionized state can only be reached directly if ICD occurred before the arrival of the probe pulse. Otherwise, without ICD, a $\text{Ne}^{+}- \text{Ne}^{+}$ or $\text{Ne}^{2+}- \text{Ne}$ pair is created. Hence, the yield of coincident $\text{Ne}^{2+} - \text{Ne}^{+}$ ions is expected to increase with $t_D$ at a slope that reflects the ICD lifetime.

\begin{figure}
\includegraphics[width=0.97\columnwidth]{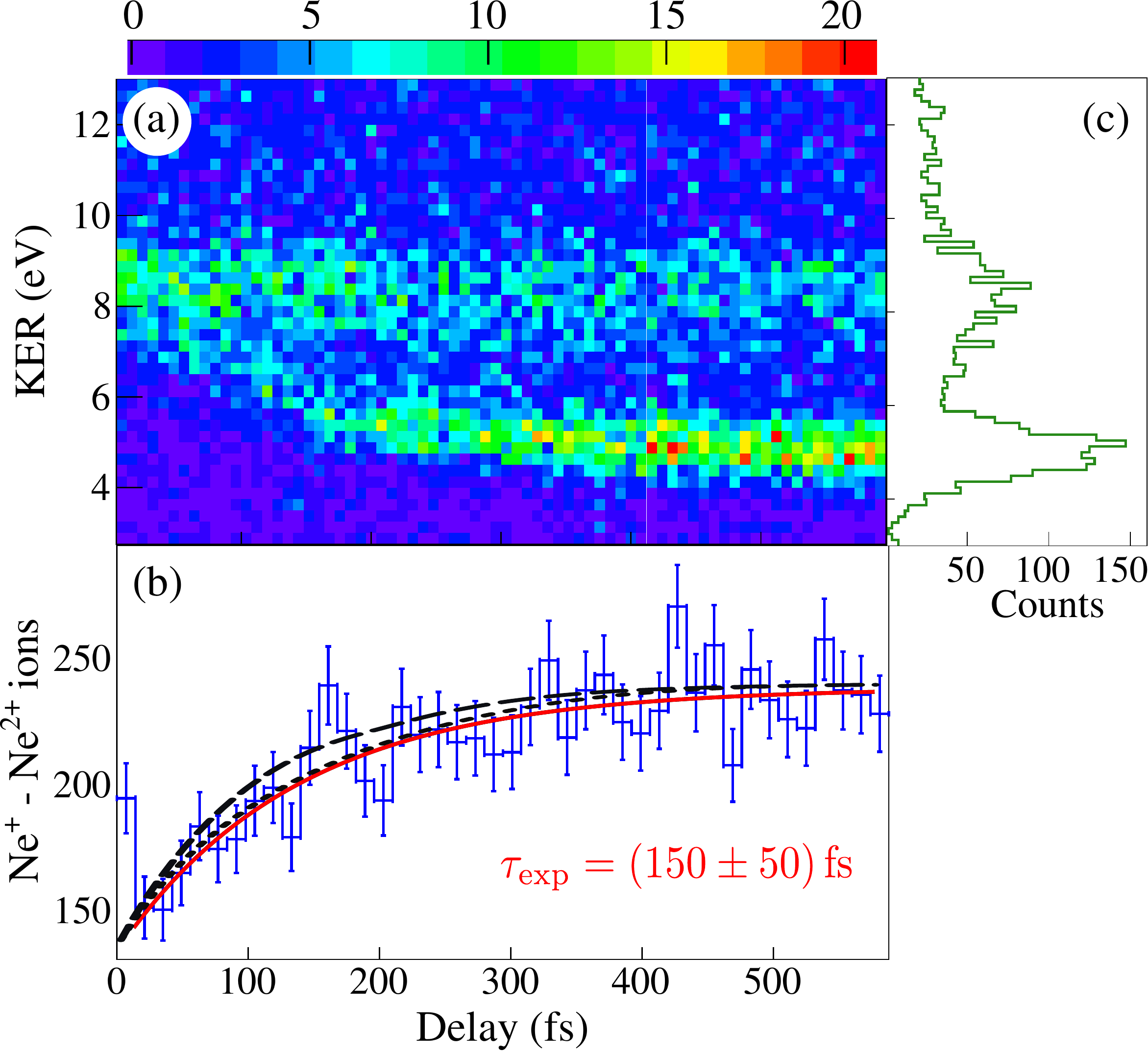}
\caption{\label{fig:data} (a) KER of coincident $\text{Ne}^{+} - \text{Ne}^{2+}$ fragments as a function of pump-probe delay $t_D$. (b) Projection of all KERs between ${3}$ and $\SI{11}{\electronvolt}$ onto the delay axis. The ICD lifetime is determined via an exponential fit (red line). For comparison, we show quantum mechanical calculations (black line) including nuclear dynamics for the wave packet evolving on the $2^2\Sigma^+_u$ (long-dashed lines) or $2^2\Sigma^+_g$ (short-dashed lines) state of $\text{Ne}^{+}_2(2s^{-1})$. The theoretical curves are scaled according to the experimental statistics. (c) Projection onto the KER axis for $t_D$ between ${400}$ and $\SI{600}{\fs}$. 
}
\end{figure}

In Fig.\,\ref{fig:data}(a) the measured KER of coincident $\text{Ne}^{2+} - \text{Ne}^{+}$ pairs is plotted as a function of $t_D$. The spectrum shows two clear features: a time-dependent one with decreasing KER towards larger $t_D$ and a delay-independent contribution for KERs larger than $\SI{7}{\electronvolt}$. The latter originates from direct multiphoton absorption during a single pulse from the ground state into a repulsive $\text{Ne}^{2+} - \text{Ne}^{+}$ state. The projection of KERs at large $t_D$, as shown in Fig.\,\ref{fig:data}(c), peaks around $\SI{8.5}{\electronvolt}$. This corresponds to the Coulomb energy of one singly and one doubly charged ion at a distance of $3.2 \einh{\AA}$, and thus confirms single-pulse multiphoton ionization from the ground-state equilibrium distance $R_{\text{eq}} = \SI{3.1}{\angstrom}$. 
The time-dependent component exhibits a clear signature of delayed ionization. At small $t_D$, the KER matches that of the time-independent process, confirming a rapid population of the final charge state. However, towards larger $t_D$ the KER approaches the Coulomb energy of two singly charged ions separated by $R_{\text{eq}}$. In this signal the information about the ICD lifetime is contained. To resolve the $\text{Ne}^{+} - \text{Ne}^{2+}$ channel at delays close to zero it is necessary to reduce the contribution of multiphoton ionization, as we have done by maintaining a rather low FEL intensity.

\begin{figure}
\includegraphics[scale=0.91]{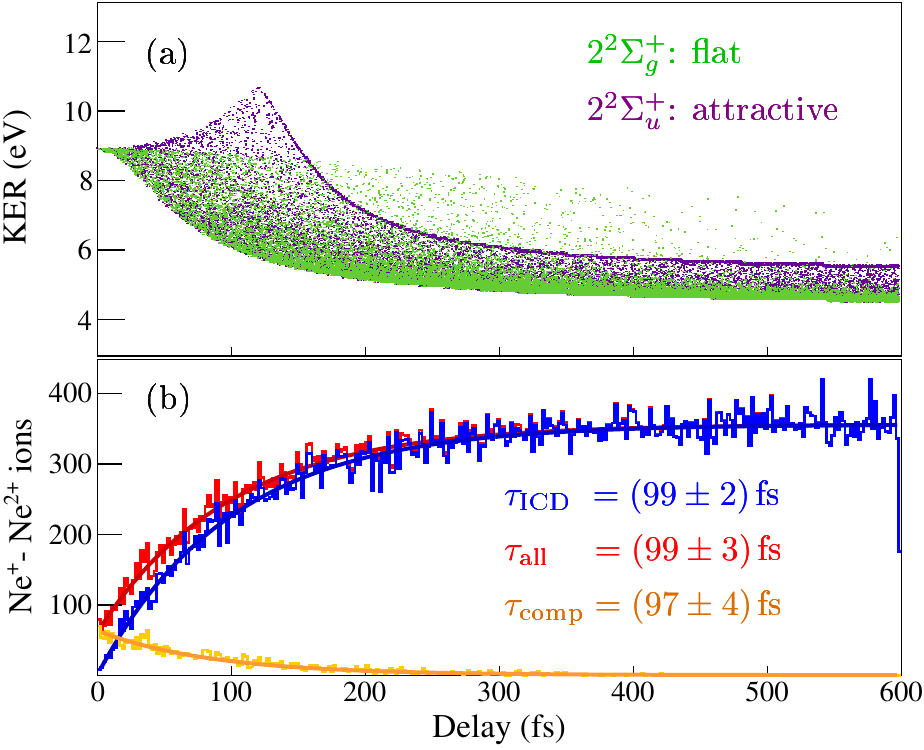}
\caption{\label{fig:simulation} (a) Simulation of KER versus time delay for the primary ICD pathway into the fragmentation channel $\text{Ne}^{2+} - \text{Ne}^{+}$ with an input lifetime of $\SI{100}{\fs}$. Propagation on the $2^2 \Sigma _{g}^{+}$ state is drawn in green while the purple color denotes the $2^2 \Sigma _{u}^{+}$ state. (b) Projections of the simulated KER versus time delay spectra on the delay axis for the primary ICD (blue line) and the competing $\text{ICD}_2$ (orange line). The sum of both is shown in red. Solid lines are exponential fits to the data.} 
\end{figure}

We now address the question of how the ICD lifetime is contained in the low-energy trace of Fig.\,\ref{fig:data}(a). To answer it, we perform a classical simulation of the pump-probe experiment in which the nuclear motion is modeled by a pointlike particle moving on the corresponding PECs. According to the Franck-Condon approximation the pump process is modeled by placing the particle in the intermediate $2^2 \Sigma _{g}^{+}$ or  $2^2 \Sigma _{u}^{+}$ state at $R_{\text{eq}}$. The subsequent nuclear motion is calculated by solving Newton's equation of motion along the internuclear distances $R$. The ICD of the intermediate state is implemented by assuming an exponential decay probability $\exp({-{t_{\text{ICD}}}/{\tau}})$ with a constant lifetime $\tau$. After the decay the particle pursues its motion on the Coulomb potential curve $1/R$, which is a good approximation of the molecular curve, in particular, at large $R$, until the delayed probe pulse brings it onto the steeper $2/R$ curve of the $\text{Ne}^{2+} - \text{Ne}^{+}$ fragmentation channel. The KER for a specific $t_D$ is obtained by adding up the kinetic energies that the particle has gained during its motion on the individual PECs. 

The simulated KER versus delay spectrum for $\text{Ne}^{2+} - \text{Ne}^{+}$ ion pairs is shown in Fig.\,\ref{fig:simulation}(a) for an exemplary ICD lifetime of ${\tau}=\SI{100}{\fs}$. The general trend of decreasing KER towards longer delays known from the experimental data is well reproduced. The decrease is a consequence of the different steepnesses of the molecular (or Coulombic) potential curves for the doubly and triply charged dimer: The faster the system is promoted to the steeper 2/$R$ curve, the smaller the internuclear distance, because the nuclei do not have much time to move apart. This results in large KERs for short time delays and vice versa. 

The spread of simulated KERs at a given delay is due to the statistically varying time when ICD occurs: It may happen at any time $t_{\text{ICD}}$ between the two pulses. ICD immediately after the pump pulse means little nuclear motion on the inner-valence potential ($2^2 \Sigma _{g}^{+}$ or $2^2 \Sigma _{u}^{+}$) and a rather long time for dissociation along the $1/R$ Coulomb curve. This results in events close to the low-KER cutoff of Fig.~\ref{fig:simulation}(a). In contrast, if ICD happens late but just before the arrival of the probe pulse, higher KERs are reached. In this case enough time is left for nuclear motion, which sensitively depends on the shapes of the $2^2 \Sigma _{g}^{+}$ and $2^2 \Sigma _{u}^{+}$ curves. While nuclear motion on the virtually flat $2^2 \Sigma _{g}^{+}$ potential is negligible [green in Fig.\,\ref{fig:simulation}(a)], the attractive $2^2 \Sigma _{u}^{+}$ state [purple in Fig.\,\ref{fig:simulation}(a)] leads to an increase of the KER for $t_{\text{ICD}}$ close to $t_D$ and delays smaller than $\approx \SI{120}{\fs}$, because the internuclear distance has decreased before the decay. Since ICD is energetically allowed as long as the $\text{Ne}^{+} (2p^{-1}) - \text{Ne}^{+} (2p^{-1})$ PEC lies below the $\text{Ne}^{+} (2s^{-1}) - \text{Ne}$ curve, no KERs higher than those resulting from a Coulomb explosion at the curves' crossing point around $R=\SI{2.6}{\angstrom}$ can occur. For the small fraction of $\text{Ne}^{+}_{2}$ ions in the $2^2 \Sigma _{u}^{+}$ state that reach the crossing point, we assume immediate decay at $R=\SI{2.6}{\angstrom}$ onto the $\text{Ne}^{+} (2p^{-1}) - \text{Ne}^{+} (2p^{-1})$ potential curve on which the particle pursues its motion until the probe pulse arrives.

The general behavior of the experimental data [Fig.\,\ref{fig:data}(a)] is well reproduced by our classical simulation supporting the proposed pathway. The increase of KER due to nuclear motion in the attractive $2^2 \Sigma _{u}^{+}$ state is not resolved in the measurement, most likely because of limited statistics and contributions from multiphoton transitions, leading to similar KERs.  

Utilizing our simulation we want to discuss the influence of competing pathways that also result in coincident $\text{Ne}^{+} - \text{Ne}^{2+}$ fragments. Until now we have only considered the case where ICD occurs in the time interval between pump and probe pulse. An additional contribution arises when the probe pulse arrives while the system is still in the intermediate $2^2 \Sigma _{g}^{+}$ or $2^2 \Sigma _{u}^{+}$ state. Since a single photon of $\SI{58.2}{\electronvolt}$ cannot further ionize both sides of the $\text{Ne}_{2}^{+}$ molecular ion, the probe pulse will either remove a $2s$ or a $2p$ electron from the neutral $\text{Ne}$ atom, or create a further $2p$ vacancy in the $\text{Ne}^{+}$ ion. In total four different states can be accessed, which are weighted according to their multiplicity. One particular state, $\text{Ne}^{2+} (2s^{-1}2p^{-1}) - \text{Ne}$, populated by probe $B$ in Fig.\ref{fig:potential}, can itself undergo ICD into $\text{Ne}^{+} - \text{Ne}^{2+}$ fragments ($\text{ICD}_2$). 

In order to investigate the influence of $\text{ICD}_2$ on the total yield of $\text{Ne}^{+} - \text{Ne}^{2+}$ ions, we apply our classical simulation to this pathway. Compared to the primary ICD we assume a population probability of 1/6 for $\text{ICD}_2$, due to multiplicity considerations and neglecting $2s$ ionization because of the low cross section. This is a conservative estimate because it does not take other possible decay mechanisms of the $\text{Ne}^{2+} (2s^{-1}2p^{-1}) - \text{Ne}$ state into account \cite{Stoychev08_2}.  

In Fig.\,\ref{fig:simulation}(b) the time-dependent yield of $\text{Ne}^{+} - \text{Ne}^{2+}$ ions is plotted for the primary channel, the competing channel, and the sum of both, integrated over all KERs. An exponential fit to the slope of the yield for the primary pathway reproduces the simulation's input ICD lifetime of $\approx100 \einh{fs}$ [blue curve in Fig.\,\ref{fig:simulation}(b)]. The behavior of the ion yield for this channel is intuitively clear if we assume that the $\text{Ne}^{2+} - \text{Ne}^{+}$ states can only be accessed after ICD has occurred. Thus, we probe the number of $\text{Ne}_{2}^{+}$ ions that have undergone ICD as a function of time, yielding a measure of the lifetime. As one would expect, the $\text{Ne}_{2}^{+}$ ICD lifetime can also be extracted from the competing process $\text{ICD}_2$. It displays a signal that is exponentially decreasing with $t_{D}$ [orange curve in Fig.\,\ref{fig:simulation}(b)] since it is dominant for short delays when only few $\text{Ne}_{2}^{+}$ ions have undergone ICD. The summed spectrum -- which corresponds to our measured data -- follows an exponential rise, still reproducing the input lifetime $\tau$ within error bars. 

The discussed $\text{Ne}^{2+} (2s^{-1}2p^{-1}) - \text{Ne}$ state leading to $\text{ICD}_{2}$ can also be accessed by direct two-photon absorption from the ground state. This component causes a constant background of around $\SI{8.5}{\electronvolt}$ for $\text{ICD}_{2}$, overlapping with direct multiphoton absorption at $R_{\text{eq}}$, both do not influence the lifetime fit. Furthermore, $\text{ICD}_{2}$ does not alter the delay dependence of the $\text{Ne}^{+} - \text{Ne}^{2+}$ yield as long as the $\text{ICD}_{2}$ lifetime is much shorter than the radiative lifetime.    

In summary, the simulation shows that the delay-dependent ion yield enables us to robustly deduce the ICD lifetime of the $2s$ inner-valence state in $\text{Ne}_{2}^{+}$, irrespective of other pathways that lead to the same final state. Thus, we project the measured KER versus delay distribution for KERs between ${3}$ and $\SI{11}{\electronvolt}$ onto the delay axis [Fig.\,\ref{fig:data}(b)]. From the exponential fit to the experimental data [Fig.\,\ref{fig:data}(b)] we extract an ICD lifetime of $\SI{150(50)}{\femto\second}$. 

The given error is purely statistical and larger than the uncertainty of $\pm\SI{36}{\femto\second}$ resulting from a FWHM average pulse duration of \SI{60}{\femto\second}. However, an even better temporal resolution is expected in pump-probe experiments due to the spiky temporal structure of the FEL pulses~\cite{Meyer12}. This structure manifests itself in our data as a sharp peak around $t_{D} = 0$ [Fig.\,\ref{fig:data}(b)], revealing the coherence length of the pulse and causing an increased rate of multiphoton processes \cite{stickstoff}. For the lifetime fit we ignore this feature as it is unrelated to ICD. On the other hand, it allows us to determine the zero point of the delay axis to an accuracy of a few femtoseconds. However, since the coherence spike arises for the lowest $\text{Ne}^{+} - \text{Ne}^{2+}$ yield around $t_{D} =0$ our fit result should be understood as an upper limit for the lifetime. 

The measured lifetime cannot be compared directly to computed ICD rates at a fixed internuclear distance or at a range of such \cite{Santra01_2,fano-adc,Vaval07}. Instead, quantum mechanical calculations including the nuclear wave packet dynamics in the intermediate $2^2\Sigma^+_{g,u}$ states of $\text{Ne}_2^+$ must be performed. Previous studies, such as \cite{Scheit04}, concentrated on the prediction of the ICD electron and KER spectra. Here, we present theoretical results where we focus on the norm of the nuclear wave packet evolving on one of the intermediate PECs of $\text{Ne}_2^+$ (Fig.\,\ref{fig:potential}) using the computational technique of \cite{Scheit03}. We start from the neutral dimer in its ground electronic and vibrational state and use a set of $R$-dependent decay widths $\Gamma(R)$: Fano-algebraic diagrammatic construction widths of \cite{fano-adc} [$\Gamma(R)$ of the $2^2\Sigma^+_g$ state, not reported in \cite{fano-adc}, was calculated by a procedure fully analogous to the one used for the $2^2\Sigma^+_u$ state]. 

In addition to the lifetime fit, Fig.\,\ref{fig:data}(b) shows the comparison of measured $\text{Ne}^{+} - \text{Ne}^{2+}$ pairs as a function of time with results of the quantum calculation using the $R$-dependent decay widths. The decay dynamics of the gerade and the ungerade states is not dramatically different, the difference being comparable to the experimental uncertainty in the decay rate. Clearly, the experimental results compare well with the quantum calculations including nuclear dynamics [black curves in Fig.\,\ref{fig:data}(b)]; thus, the present experiment allows us to test directly the theoretical predictions of the $R$-dependent ICD width. 

The presented single-photon pump single-photon probe approach is a universal method to determine lifetimes of ICD in small clusters. In the present study we already discussed the population of an additional excited state $\text{Ne}^{2+} (2s^{-1}2p^{-1})- \text{Ne}$ undergoing ICD whose lifetime could be determined in the same fashion with different pulse parameters. Additionally, the investigation of NeAr is an interesting example where the dynamics of a single excited state can be studied \cite{Scheit06} instead of the superposition of $2^2 \Sigma _{g}^{+}$ and $2^2 \Sigma _{u}^{+}$ as in the present experiment.

\begin{acknowledgements}
We acknowledge technical support from B. Knape and C. Kaiser, and fruitful discussion with Y.C. Chiang. The authors thank the scientific and technical team at FLASH for optimal beam time conditions. Support from the ASG at CFEL and the DFG via Grant No. Kl-1439/5 is gratefully acknowledged. A.R. and M.F.K. were supported by the Chemical Sciences, Geosciences, and Biosciences Division, Office of Basic Energy Sciences, Office of Science, U.S. Department of Energy under Grant No. DE-FG02-86ER13491. S.M. is grateful to JSPS for financial support. K.U. acknowledges support for X-ray Free Electron Laser Utilization Research Project and the X-ray Free Electron Laser Priority Strategy Program by MEXT. S.S. is grateful to the Theoretical Chemistry group of the University of Heidelberg for permission to use its computer cluster to run the calculations. V.A. is supported by the EPSRC (UK) through the Career Acceleration Fellowship and the Programme Grant on Attosecond Dynamics. Y.H.J. acknowledges support from the National Basic Research Program of China (973 Program)(Grant No. 2013CB922200), the National Natural Science Foundation of China (grant 11274232), and Hundred Talents Program of the CAS.

\end{acknowledgements}

\bibliography{schnorr}

\end{document}